\documentclass[fleqn,usenatbib]{mnras}

\usepackage{newtxtext,newtxmath}

\usepackage[T1]{fontenc}

\DeclareRobustCommand{\VAN}[3]{#2}
\let\VANthebibliography\thebibliography
\def\thebibliography{\DeclareRobustCommand{\VAN}[3]{##3}\VANthebibliography}

\usepackage{graphicx}	
\usepackage{amsmath}	
\usepackage{subfiles}
\usepackage{xcolor}
\usepackage{hyperref}
\usepackage{orcidlink}

\usepackage{siunitx}
\DeclareSIUnit\year{yr}
\DeclareSIUnit\parsec{pc}
\DeclareSIUnit\msun{M_\odot}
\DeclareSIUnit\Rsun{R_\odot}

\newcommand{\kms}{\unit{\km\per\s}}
\newcommand{\kpc}{\unit{\kilo\parsec}}

\defcitealias{petersenDetectionMilkyWay2021a}{PP21}
\defcitealias{yaaqibRadialVariationLMCinduced2024a}{YPP24}

\newcommand{\ignore}[1]{}

\usepackage{pifont}
\usepackage{soul}

\title[	
Constraining the MW halo with reflex motion]{Constraining the Milky Way Dark Matter halo with LMC-induced reflex motion} 
\author[Yaaqib, Petersen, Pe\~{n}arrubia]{
Rashid Yaaqib$^{1,2}$\thanks{E-mail: rashid.yaaqib@ed.ac.uk}\orcidlink{0009-0003-9063-1382},
Michael S. Petersen $^{1}$\orcidlink{0000-0003-1517-3935},
Jorge Pe\~{n}arrubia$^{1,3}$\orcidlink{0009-0004-0611-7573}
\\
$^{1}$Institute for Astronomy, University of Edinburgh, Royal Observatory, Blackford Hill, Edinburgh EH9 3HJ, UK\\
$^{2}$Department of Physics, United Arab Emirates University, Al Ain, Abu Dhabi, UAE\\
$^{3}$Centre for Statistics, University of Edinburgh, School of Mathematics, Edinburgh EH9 3FD, UK \\
}

\date{Accepted XXX. Received YYY; in original form ZZZ}

\pubyear{2024}

\begin{document}

\label{firstpage}
\pagerange{\pageref{firstpage}--\pageref{lastpage}}
\maketitle

\begin{abstract}
Modelling perturbations of the Milky Way (MW) halo induced by the infall of the Large Magellanic Cloud (LMC) offers new avenues to constrain the dark matter (DM) distribution in our Galaxy. A key observable is the reflex motion of the Milky Way disc with respect to the halo induced by the LMC’s infall, which imprints a velocity dipole on kinematics of halo stars. Here we investigate how the dipole varies with Galactocentric radius, and study the sensitivity of the reflex motion signal to different DM outer-halo profiles. 
Using a suite of basis function expansion (BFE) simulations with truncated NFW profiles ($\rho \propto r^{-\beta}$ beyond $r=50$ kpc),
our $N$-body models show that (i) The reflex motion amplitude varies with Galactocentric radius but is largely insensitive to the outer DM slope, implying that the MW–LMC mass ratio alone does not set the dipole strength.
(ii) In contrast, the direction of the disc motion is very sensitive to the density distribution of the outer DM halo.
(iii) The contraction of the MW halo induced by the LMC's gravitational pull also depends strongly on the outer DM halo profile.
 (iv) We find a halo instability whose oscillation frequency increases with $\beta$ producing a potentially observable signature - a sinusoidal pattern of the mean radial velocity of halo stars. Finally, using BFE coefficients we find that steeper truncations  produce smaller dipole distortions, while amplifying the quadrupole distortion. These results highlight the limited constraining power of the reflex motion amplitude alone for outer MW profile parameters.

\end{abstract}

\begin{keywords}
Galaxy: general; Galaxy: kinematics and dynamics; Galaxy: halo
\end{keywords}


\section{Introduction}
\label{sec:intro}
The presence of the Large Magellanic Cloud (LMC) in the Milky Way (MW) halo has prompted much research into the dynamical response of the MW halo to the infall of the LMC. Over the past decade, numerical simulations of the LMC's infall have found that the MW disc's barycentre is displaced from the halo barycentre \citep{gomezITMOVESDANGERS2015a, petersenReflexMotionMilky2020b}. This displacement occurs in cases where the LMC's mass is $>\sim10\%$ of the MW's mass. Coupling the aforementioned findings and recent independent measurements place the mass of the LMC that place it at $1.0-2.5\times10^{11}~\rm{M}_{\odot}$ \citep{penarrubiaTimingConstraintTotal2015, erkalTotalMassLarge2019a, vasilievTangoThreeSagittarius2020, shippMeasuringMassLarge2021a, correamagnusMeasuringMilkyWay2021, koposovS5ProbingMilky2023, vasilievEffectLMCMilky2023b}. Many signatures of this displacement have been detected in the kinematics of the MW halo stars. \citet{erkalDetectionLMCinducedSloshing2021} find a north-south asymmetry, while \citet[][PP21 hereafter]{petersenDetectionMilkyWay2021a} model the velocity field of the MW halo as a dipole, where the orientation of the dipole points along the past trajectory of the LMC. This disc-displacement induced asymmetry in the outer halo was also measured in recent DESI data \citep{bystromExploringInteractionMW2024}.
This displacement leads to a apparent velocity dipole that is observed in the outer halo stars of the MW \citep{petersenDetectionMilkyWay2021a, yaaqibRadialVariationLMCinduced2024a, bystromExploringInteractionMW2024,chandraAllskyKinematicsDistant2024}. The detection of this signal calls forth a need for non-equilibrium models of the MW that account dynamical response of the MW halo the LMC. \citet{lilleengenEffectDeformingDark2022} show through self-consistent models of the MW-LMC interaction that the LMC induces a strong density dipole in the MW halo that has grown in strength over the last $\sim500~\rm{Myr}$. 

To date, few direct links between the reflex motion and the parameters of the pre-infall MW halo (or LMC halo) are known. Only a small fraction of the possible combinations between MW and LMC mass profiles prior to infall have been explored. In \citetalias{petersenDetectionMilkyWay2021a}, it was demonstrated clearly the relationship between the magnitude of reflex motion and the assumed mass of the LMC (See the Supplementary Data section). In their experiment, by keeping the LMC on a fixed trajectory and varying the mass, the resulting effect was that the travel velocity of the disc has increased. Hence, their work showed that the disc velocity has strong dependence on the LMC's assumed mass. In an extension to the work in \citetalias{petersenDetectionMilkyWay2021a}, \citet[][YPP24 hereafter]{yaaqibRadialVariationLMCinduced2024a} measured the reflex motion as a function of galactocentric distance in the data and also in a host of literature simulations (See their Table 1 for a summary of the simulations analysed). In their work, they show the utility of using the radial variation of reflex motion to compare with reflex parameters computed directly from simulations. While a single measurement of reflex motion against the outer halo shows the dislodging of the MW disc from the previously shared barycentre with the outer MW halo, it does not provide much constraining power for the profile of the MW. Measuring the radial variation can provide constraints on MW-LMC simulations as different pre-infall mass profiles of the MW will have unique responses to the infall of the LMC, leading to a different reflex motion signature. 

The goal of this work is to address how the (observable) reflex motion parameters vary in response to changes in the initial density distribution of the MW DM halo. Previous simulations show that the reflex motion parameters (magnitude and apex location) as a function of galactocentric distance owing to the differing assumptions about both the MW and LMC pre-infall parameters, such as the LMC-MW mass ratio and the scale radii used. Also, these simulations used different initial conditions of the LMC - resulting in different trajectories. Since the direction of disc motion w.r.t to the outer halo points at a location along the past trajectory of the LMC, the choice of the LMC initial position and velocity also affects the resulting reflex motion signature.
The wide array of choices made for the MW and LMC undoubtedly results in a range of predictions for the apex location, magnitudes of the disc velocity, and bulk motions in the outer stellar halo, which show us the collective response of the outer Galaxy to the LMC infall --once the disc reflex motion has been removed. To motivate this paper, we aim to explore the response of the MW stellar halo to the LMC's infall, and analyse the how to reflex motion varies as a function of the initial MW dark matter(DM) halo profile. To this aim, in this paper we present and analyse a suite of controlled self-consistent $n$-body simulations constructed with MW DM halo profiles with a truncation in the density profile. We hope to gain understanding of the interplay between the observable reflex motion parameters and the pre-infall configuration of the MW (and LMC). As will be shown in later sections, our simulations cover a wide range of MW-LMC mass profiles (and therefore mass ratios) that help address this.

This paper is structured as follows; in section \ref{sec:methods} we show the details of the density model used, and summarise the simulation parameters used. In section \ref{subsec:Numerical-methods} we present extended details on the construction of the models in the basis function expansion framework, and discuss the initial condition generation methodology. We also summarise the methodology used to calculate the reflex motion parameters in \ref{subsec:ReflexMotionParameterCalculation}. In section \ref{sec:results} we present the measurements of the reflex motion parameters in all simulations. Section \ref{sec:Discussion} discusses the results in the context of the collective response of the halo kinematics. Finally in section \ref{sec:Conclusion} we conclude the main results and insights of this work.

\section{Methods}
\label{sec:methods}

    \begin{figure}
        \includegraphics[width=\columnwidth]{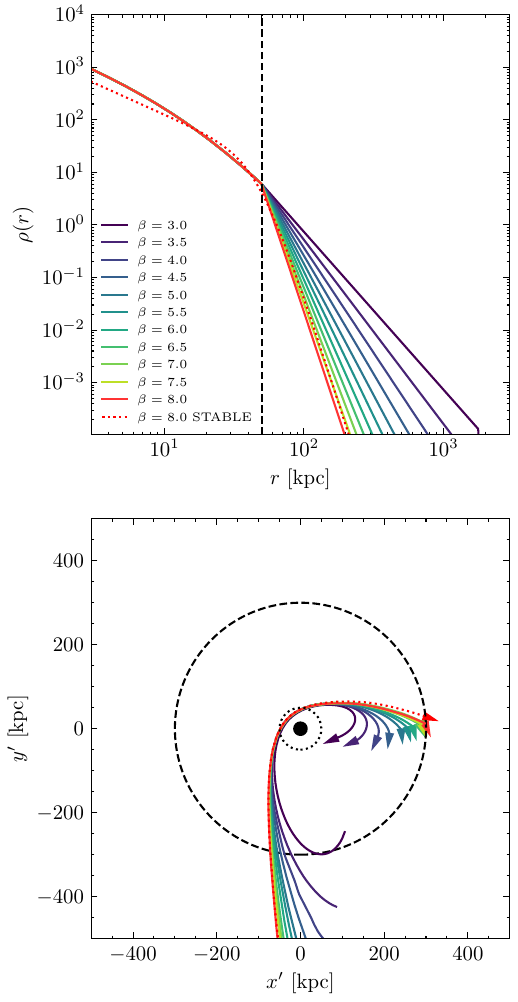}
        \caption{\textit{Top Panel:} The truncated NFW profiles of the MW haloes used in this work. The density reported in the figure is in dimensionless units. The most truncated halo has an outer halo of $\beta=8$ and the no truncation NFW continues across the break with $\beta=3$. \textit{Bottom Panel:} The trajectories of the softened Plummer sphere LMC haloes from {\sc EXP} simulations in the orbital plane of the LMC. The black point marks the centre of the MW disc. The black dotted line marks the truncation radius of 50 kpc, and the dashed line marks 300 kpc, which is the virial radius of the $\beta=3$ model. }
    
        \label{fig:cuspy-profiles}
    \end{figure}

        \begin{table*}
        \caption{\label{tab:models} Summary of MW models used in this work.}
        \begin{tabular}{ccccccc}
            \hline
            Simulation & $M_{\rm MW}$ & $M_{\rm LMC}/M_{\rm MW}$ & Outer Slope $\beta$  & $\vec{x}_{{\rm LMC}\to{\rm MW}}$ & $\vec{v}_{{\rm LMC}\to{\rm MW}}$ \\
            Unit & $ \times10^{12} M_{\odot}$ & & & $\rm{kpc}$ & $\rm{kms^{-1}}$\\
            \hline
            CU0 & 1.624 & 0.095 & 3.0 & $[11.3, 251.0, 108.7]$ & $[11.8, 72.7, -23.3]$ \\
            CU1 & 0.952 & 0.162 & 3.5 & $[34.4, 431.5, 91.2]$ & $[3.9, -13.8, -34.8]$ \\
            CU2 & 0.721 & 0.215 & 4.0 & $[48.2, 526.4, 71.3]$ & $[-0.6, -53.2, -34.3]$ \\
            CU3 & 0.604 & 0.257 & 4.5 & $[58.3, 589.4, 52.0]$ & $[-3.8, -73.6, -29.9]$ \\
            CU4 & 0.544 & 0.285 & 5.0 & $[66.2, 634.8, 34.0]$ & $[-5.8, -88.6, -26.4]$ \\
            CU5 & 0.511 & 0.303 & 5.5 & $[70.6, 660.9, 24.8]$ & $[-7.1, -96.9, -24.3]$ \\
            CU6 & 0.487 & 0.318 & 6.0 & $[75.1, 684.8, 13.3]$ & $[-8.3, -104.2, -22.1]$ \\
            CU7 & 0.469 & 0.330 & 6.5 & $[79.1, 704.7,2.2]$   & $[-9.3, -110.2, -19.9]$ \\
            CU8 & 0.457 & 0.339 & 7.0 & $[81.5, 717.5, -4.1]$ & $[-9.0, -114.0, -18.5]$ \\
            CU9 & 0.448 & 0.346 & 7.5 & $[83.3, 727.4, -8.9]$ & $[-10.4, -116.9, -17.5]$ \\
            CU10 & 0.439 & 0.353 & 8.0 & $[85.9, 738.9, -16.7]$ & $[-11.1, -120.3, -15.9]$ \\
            CU10S&  0.445& 0.337 & $8.2^{\rm a}$ &$[85.3, 736.5, -15.1] $   &  $[-10.9, -119.4, -16.2]$    &                         \\
            \hline
        \multicolumn{6}{l}{$^{\rm a}$The outer slope of the stable ABG model is slightly larger as we use non-integer values of $\beta$ when matching the model to the CU10 truncated NFW model.}\\
        \end{tabular}
    \end{table*}

\subsection{Truncated halo models}
\label{subsec:analyticmodel}

To study the MW halo response while varying the profile of the MW, we utilise a modified NFW profile. 
The truncated NFW profile consists of a piecewise density distribution with a NFW profile up to some radius $r_{\rm b}$, beyond which the profile is power-law truncated as $r^{-\beta}$. The profile is spherically symmetric and due to the inclusion of the outer profile, is finite in mass for values of $\beta > 3$ (ref).
\begin{equation}
\label{eq:truncmodel}
    \rho(r) = \rho_0
    \begin{cases}
        \frac{1}{(\frac{r}{r_s})(1 + \frac{r}{r_s})^2 }, & r < r_b \\
        \frac{1}{(\frac{r_{\rm b}}{r_s})(1 + \frac{r_b}{r_s})^2} (\frac{r}{r_{\rm b}})^{-\beta}, & r > r_{\rm b}
    \end{cases}
\end{equation}

Where the prefactor in the $r>r_{\rm b}$ region of the profile is the normalisation to the density such that the density is continuous across the break radius. The prefactor was computed by requiring that the density at the break radius are equal in the two regions.
The top panel of Figure \ref{fig:cuspy-profiles} shows the 12 models we use in this work. The models span a range of outer slopes of $3.0 < \beta < 8.0$ with a break radius of $r=50~\rm{kpc}$. For models with $\beta=[3.0,3.5]$ an error function truncation was applied at $r= 600~\rm{kpc}$ to ensure the models are compatible with the simulation software inputs. As can be seen from the figure, all the models share the same density profile interior to the break radius. This behaviour was set by construction to limit our experiment to changes in the outer slope only. The mass interior to the break radius was set to $M=0.4\times10^{12}~\rm{M}_\odot$ for all models, consistent with observations of the MW \citep{mcmillanMassDistributionGravitational2016, bland-hawthornGalaxyContextStructural2016a}. The differences in mass arise solely from the outer slope variation. Table~\ref{tab:models} lists the profile parameters used for all 12 simulations.

While the density profile is continuous, its first derivative is discontinuous at $r=r_{\rm b}$. This feature in the profile could lead to the formation of instabilities in the models when evolved in isolation. In order to facilitate the interpretation of results in later sections in the presence of an instability, we include an additional simulation of the $\beta=8.0$. The stable simulation is an $(\alpha,\beta,\gamma)$ (ABG) profile \citep{1996Zhao} with parameters $(3.6,8.2,1.2)$ with a scale radius of $r_s=48~\rm{kpc}$ (See Equation~\ref{eq:ABGmodel} for the density profile of this model).
\begin{equation}
\label{eq:ABGmodel}
    \rho(r) = \rho_0 \, \left( \frac{r}{r_s} \right)^{-\gamma} \left[ 1 + \left( \frac{r}{r_s} \right)^{\alpha} \right]^{\frac{\gamma-\beta}{\alpha}}  \,.
\end{equation}

The ABG profile was constructed by simple optimisation of the ABG profile parameters such that they match the $\beta=8.0$ model. This ABG model also contains the same interior to the break radius as the aforementioned models. This model is a taken to be a stable counterpart of the truncated model, where the stability was checked using Equations 5,6 in \citet{dattathriDipoleInstabilityGravitational2025}. The density profile of this model is shown as the dotted line in Figure~\ref{fig:cuspy-profiles}. The figure shows small deviations between the ABG and truncated NFW models in the interior regions ($r<10~\rm{kpc}$).

    \subsection{Numerical methods}
        \label{subsec:Numerical-methods}
        In this section we outline briefly the numerical methods used to generate particles in each simulation and the method of finding the initial conditions of the LMC-MW using reverse-time integration. We also summarise the reflex motion parameters used throughout the paper.

\subsubsection{Simulation initialisations}
All simulations and MW disc, halo and LMC halo are modelled using the basis function expansion (BFE) code {\sc exp~}\citep{petersenEXPPythonPackage2025}.\footnote{\hyperlink{https://github.com/exp-code}{https://github.com/exp-code}} For each MW halo model, the model consists of an exponential disc and a truncated NFW halo initially. The disc initialisations are identical in all simulations consisting of $10^{6}$ particles and modelled as an exponential disc. All remaining disc initialisation parameters are kept as the default parameters of the {\tt gendisk} tool provided with {\sc exp}. For the MW halos there are $10^7$ particles and for a given value of $\beta$ we tabulate the density, potential and enclosed mass of each model and use {\tt gendisk} for particle initialisation. All the {\sc exp} parameters that were used are summarised in Table~\ref{tab:model-summary}.\footnote{Note that only the modified parameters are listed, the remaining parameters use the default parameters present in the software.}
    
\subsubsection{Satellite initial conditions}
The initial conditions of the LMC-MW system are found using the reverse time integration method used in \citep{vasilievDearMagellanicClouds2023, vasilievTangoThreeSagittarius2020}. The technique evolves the centre of mass motion of the MW-LMC system in an inertial frame via integration of a 12 dimensional system of ordinary differential equations that evolve the phase space of the MW and LMC. During the initial condition finding, the MW and LMC are modelled using rigid non-deforming potentials. For the purposes of our work here, we do not require exact matching between the observed LMC coordinates and simulated ones; therefore this approximate trajectory-finding method provides an efficient and robust initialisation.

\subsection{Reflex motion parameters}
\label{subsec:ReflexMotionParameterCalculation}
`Reflex motion' is defined as the motion of the MW disc w.r.t. to a given shell in the halo.\footnote{Each shell is chosen to be centred on the disc barycentre.} The parameters of the reflex motion are as defined using the dipole model originally presented in \citetalias{petersenDetectionMilkyWay2021a}. The motion of the disc is defined by the $\vec{v}_{\rm travel}$ which has magnitude $v_{\rm travel}$ and directions $\ell_{\rm apex},b_{\rm apex}$ defined in the galactic coordinate system. In addition to reflex, the bulk motions are defined as the mean motions in a given shell of the halo {\it after accounting for reflex motion}. Accounting for reflex motion is critical to separate the measurement of the internal kinematics in the MW halo, and the velocity dipole signal from the motion of the disc. When not accounting for reflex, the dipole signal aliases into the mean motions of the halo. The effects are most significant on the polar and radial velocities. Since the disc motion is pointed (in general) towards the $b=-90^\circ$ galactic pole, the mean velocity in a given halo shell at larger radii appear to be moving towards the galactic north pole (resulting in positive $v_\theta$). In the mean radial motion, the amount of motion is underestimated when leaving reflex unaccounted for. Where the bulk motion parameters are the mean spherical velocities $(\langle v_r\rangle, \langle v_\phi\rangle, \langle v_\theta\rangle)$ (radial, azimuthal and polar respectively). For observations, these parameters must be fit for in a likelihood based approach (See e.g. \citetalias{petersenDetectionMilkyWay2021a}, \citetalias{yaaqibRadialVariationLMCinduced2024a}). However, in simulations they are straightforwardly approximately calculated\footnote{We note here that this is an approximation as ideally the reflex motion model of \citetalias{petersenDetectionMilkyWay2021a,yaaqibRadialVariationLMCinduced2024a} would need to be fit to fully account for the coupling between the dipole terms and bulk motion terms. Regardless the approximation agreed well with the fitting method as tested in \citetalias{yaaqibRadialVariationLMCinduced2024a}.} using the 6D phase space information of halo stars. We refer the interested reader to the measurement technique described in Section~3.2 of \citetalias{yaaqibRadialVariationLMCinduced2024a}. 

    \begin{table}
    \centering
    \caption{\label{tab:model-summary} Summary of {\sc EXP} parameters used in this work.}
    \begin{tabular}{ccccccc}
        \hline
        Parameter & Value & Notes \\
        \hline
        $N_{\rm halo}$ & $10^7$ & -  \\
        $N_{\rm disc}$ & $10^6$ & -   \\
        $M_{\rm LMC}$ & $1.55\times10^{11}~\rm{M}_{\odot}$ & -  \\
        $(l_{\rm max}, n_{\rm max})_{\rm halo}$ & (6,20) & Same for all models \\
        $(m_{\rm max}, n_{\rm max})_{\rm disc}$ & (4,10) & Same for all models \\
        $t_{\rm rewind}$ & $4.1~\rm{Gyr}$ & - \\
        \hline
    \end{tabular}
    \end{table}


\section{Results}
\label{sec:results}

\begin{figure}
    \includegraphics[width=\columnwidth]{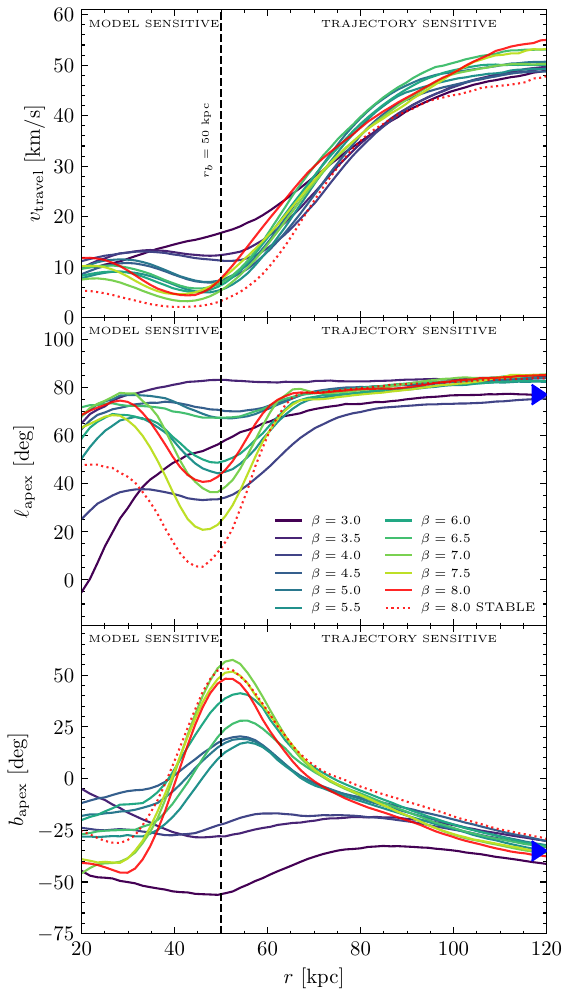}
    \caption{Magnitude and direction of the LMC-induced velocity dipole owing to the motion of the disc w.r.t to the outer halo. The reflex motion is calculated for all 12 truncated-NFW models and one stable ABG model used in this work. The inner regions of the halo are where the reflex is sensitive to the model choice (between $0-70~\rm{kpc}$). While the outer regions are sensitive to the trajectory of the LMC (between $70-120~\rm{kpc}$) respectively. The blue arrows on the right show the galactic coordinates $l_{\rm LMC},b_{\rm LMC}$ at $t-t_{\rm peri} = -400~\rm{Myr}$. The apex locations at large distances are consistent with the past location of the LMC. \textit{Top panel}: In the amplitude of disc motion, no significant differences arise when varying the outer slope; the same trend with galactocentric radius is observed for all models despite the changing (total) mass ratio. The $v_{\rm travel}$ increase remains small within $\sim 50$ kpc, but then increases almost linearly between $50-100$ before flattening at distances greater than $100$ kpc. \textit{Middle panel}: The angle $\ell_{\rm apex}$ as a function of galactocentric radius. \textit{Bottom panel}: The angle $b_{\rm apex}$ as a function of galactocentric radius.} 
    \label{fig:cuspy-reflex}
\end{figure}
\begin{figure}
    \includegraphics[width=\columnwidth]{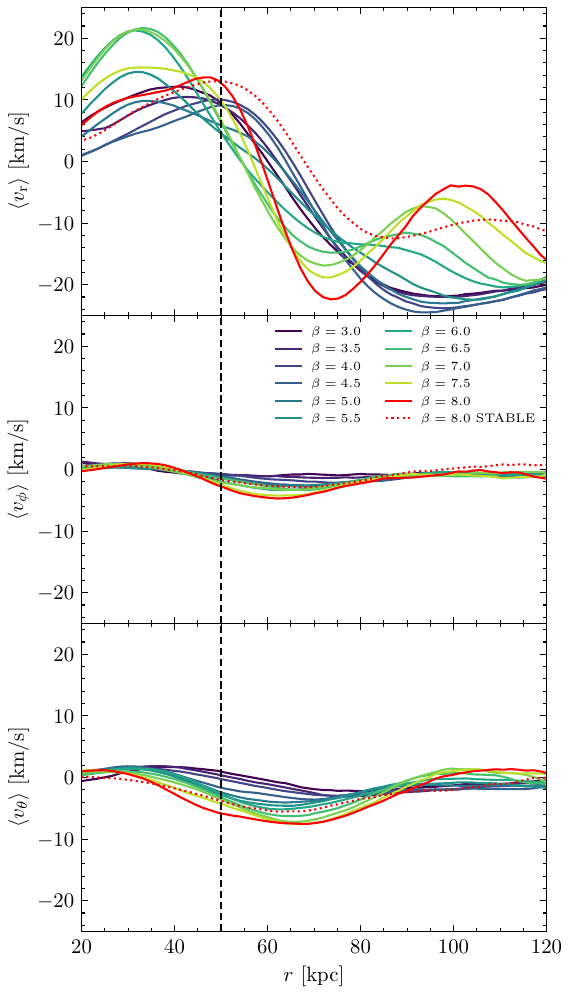}
    \caption{The reflex-corrected bulk motions in the MW haloes for each simulation at present day. \newline
    \textit{Top Panel:} The mean radial velocity in shells of width 3kpc between 20 and 120 kpc. All models show a compression signal between $\sim 50-100$ kpc. Models with sharper truncations show rebounding radial velocities at larger radii. \newline
    \textit{Middle Panel:} The mean azimuthal velocity of halo stars in all simulations. No significant trends are found with varying $\beta$, other than a mildly increasing rotation signal with decreasing outer slope. \newline
    \textit{Bottom Panel:} The mean polar velocity in each shell. }
    \label{fig:cuspy-bulkmotions}
\end{figure}

\subsection{Reflex motion}
Figure \ref{fig:cuspy-reflex} shows the reflex motion at present-day measured in the frame of the disc for all the MW haloes with an increasingly sharper truncation in the outer slope (see \ref{tab:models}). We find that increasing the outer slope of the halo does not change the magnitude of the reflex motion with distance (top panel). Despite the wide range of MW-LMC ratios probed in these simulations ($M_{\rm LMC}/M_{\rm MW}\sim 0.095-0.34$) the amplitude of the dipole varies only by about 5 kms$^{-1}$, with no clear trend with outer slope.

In the apex directions however, a clear trend is observed with increasing outer slope. 

In $\ell_{\rm apex}$ all models show a flattening in the function at distances beyond $80~\rm{kpc}$, reaching a value of $\sim 75^\circ$, pointing to the past location of the LMC at $t-t_{\rm peri}=-400~\rm{Myr}$ (See Figure~\ref{fig:pasttrajLMC} for the past trajectory of the LMC in galactic coordinates). At intermediate distances, we recover a trend with increasing outer slope in the models. The apex longitude remains relatively unchanged for models with shallow outer slopes ($4.0 < \beta < 5.0$) remaining at a value of $\sim 75^\circ$ at all radii. However, steeper outer slopes show a dip between $25-50~\rm{kpc}$, to a minimum value of $20^\circ$ for the $\beta=7.5$ model. After which the models assume by a rapid rise to $\sim 75^\circ$ by $60~\rm{kpc}$, remaining flat at larger distances and all within $10^\circ$ degrees. The $\beta=3.0$ model shows a distinct behaviour in $\ell_{\rm apex}$. It does not contain the dip in $\ell_{\rm apex}$ characteristics of models with higher values of $\beta$. It begins at a minimal value of $-5^\circ$ at $20~\rm{kpc}$, then rises rapidly to be in agreement with other models at $60~\rm{kpc}$, then asymptotes to a value of $\sim 75^\circ$ at larger distances. 

The stable $\beta=8.0$ ABG model is in good agreement with the truncated NFW counterpart in $v_{\rm travel}$ and $b_{\rm apex}$.

In $b_{\rm apex}$ we find that for models with moderate to large truncations ($\beta>4.5$), the apex direction with distances goes from being negative between $20-40~\rm{kpc}$ then positive between $40-60~\rm{kpc}$ around the truncation radius of the haloes before converging to approximately the same value at large radii for the models. At the largest radii, all simulations have similar values of $b_{\rm apex}\simeq -35^{\circ}$, which is the approximate location of the LMC $400~\rm{Myr}$ ago. At intermediate distances ($\sim 50~\kpc$), for the simulations with outer slopes between $\beta=7-8$ values of $b_{\rm apex}$ change from $-40^{\circ}$ at $20~\rm{kpc}$ to $+40^\circ$ at 50 kpc, reaching a peak value of $\sim 50^\circ$. For all models, the peaks occur at roughly the same radius, but with maximum value of $b_{\rm apex}$ decreasing with shallower truncations. The models $\beta=3.0,3.5,4.0$ show little variation of apex latitude with distance.

\subsection{Mean velocity of the stellar halo}
The MW stellar halo show strong bulk motions as a result of the LMC infall. In particular, we find clear trends in the compression signal $\langle v_r \rangle$ as a function of distance in Figure \ref{fig:cuspy-bulkmotions}. 
It is important to stress that the values of the mean velocities plotted in this Figure have been corrected for the reflex motion. Reflex motion corrections are needed to account for the aliasing of the velocity dipole signature arising from the motion of the disc, into the mean motions in the halo (discussed further in Section \ref{sec:Discussion}). 
Halo compression arises as the outer stellar halo feel the gravitational pull of the LMC on first infall, which leads to a mean radial velocity that is negative at large radii.
In all the models, the onset of compression occurs at$\sim 50~\kpc$ with the largest compression values of about $-20~\rm{kms}^{-1}$. For the NFW-like truncated model ($\beta=3.0)$ the $\langle v_r \rangle$ shows a very shallow rise to $\sim~10{\rm kms}^{-1}$ between $20-50~\rm{kpc}$ then a steep fall (compression) at larger distances. Despite the lack of truncation for this model, the distance where compression begins is consistent with models with sharper truncations. Hinting that compression starts at the approximate present-day location of the LMC, rather than being tied to the break radius.

All models show a cyclic variation in $\langle v_r \rangle$. The oscillatory behaviour of $\langle v_r \rangle$ with distance is linked to the natural instability of the models (see Section~\ref{sec:Discussion}). More truncated haloes with $\beta$ between $(5,8)$ show a bigger rebound of the compression between $80-100~\kpc$, after reaching their minimal values between $60-80~\kpc$. Generally models with shallower truncations reach their minimum value of $\langle v_r \rangle$ at larger distances.
Interior to the break radius, we find that as the steepness of the outer slope increases, more expansion is experienced by the inner halo, with the trend holding for all values of $\beta$ except the steepest two models. Finally, we find very small amplitude trends in the bulk rotation and polar motion of the halo stars. In $\langle v_\phi \rangle$ a very mild rotation signal is recovered with the amplitude of the rotation peaking around $60~\kpc$ for all models, whose value increases with the steepness of the outer slope. For the model with $\beta=-8.0$, the measure value of $\langle v_\phi \rangle$ was $\sim 5~\kms$.
\subsection{Basis function expansion coefficient amplitudes}

\begin{figure}
    \includegraphics[width=\columnwidth]{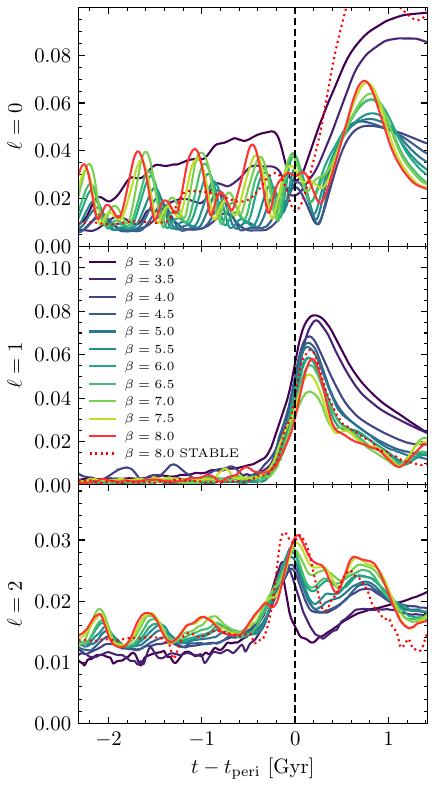}
    \caption{The relative amplitude in each of the MW halo basis functions as a function of time for each simulation. The vertical dashed line is set to the present-day time of simulations. The dotted red line shows the relative amplitudes of the stable ABG model of the $\beta=8.0$ truncated NFW model.\newline
    \textit{Top Panel:} The relative amplitude in the monopole of the basis function expansion coefficients. Oscillations in the monopole amplitude indicate the presence of an instability in the truncated NFW models. \newline
    \textit{Middle Panel:} The relative amplitude in the dipole. The dipole shows little evolution until close to present-day, as the LMC reaches pericentre the dipole amplitude increases for all models.\newline
    \textit{Bottom Panel:} The relative amplitude in the quadrupole. Models with truncations $\beta > 4.0$ show mild oscillations prior to the infall of the LMC. Close to present-day the quadrupole amplitude increases for all models.}
    \label{fig:coefficients}
\end{figure}

In this section we analyse the coefficients of the basis function expansions used to compute the potential and density of the simulation. 
These coefficients encode the variations in density and potential, and can be used to diagnose the response of the MW to the LMC -- or natural instabilities.

Figure~\ref{fig:coefficients} shows the relative amplitude in each of the monopole, dipole and quadrupole for each of the simulations. Interesting trends can be observed for all models. In the case of the monopole, models with truncations of $\beta <4.0$ show clear oscillations much before the pericentric passage of the LMC. These oscillations are indicative of a radial instability mode and has a period of $\sim 0.6~\rm{Gyr}$. The amplitude of the radial oscillations also increases with increasing $\beta$. The NFW-like models $\beta=3.0$ and $\beta=3.5$ do not show oscillations, rather, the strength in the monopole increases almost linearly for both. At present day, all models have almost equal power in $l=0$, suggesting the amplitude is tied to the passage of the LMC. Oscillations in the relative amplitude of the monopole and an increasing amplitude with time are key indicators of instability. The stable ABG $\beta=8.0$ model does not contain an instability: its evolution in $l=0$ follows that of the stable $\beta=3.0,3.5$ models. 

For the dipole, there is a clear trend with $\beta$ that we note here. At present day, as $\beta$ increases (more truncated) the dipole strength decreases. The peak amplitude in $l=1$ also follows this trend, where the largest dipole power measured was in $\beta=3.0$ and the smallest was in $\beta=7.0$. Furthermore, for the quadrupole, we find an inverse relationship. As $\beta$ increases, the quadrupolar term also increases in power at present day. We note also that periodic oscillations in $l=2$ were found for models with truncations in the past ($(t-t_{\rm peri})< 1~\rm{Gyr}$), whose amplitude also increase with increasing $\beta$. Although these oscillations are smaller in amplitude compared to the monopole. The stable ABG $\beta=8.0$ shows little evolution in both $l=1,2$ prior to passage of the LMC.

\section{Discussion}
\label{sec:Discussion}
Reflex motion is defined as the motion of the disc w.r.t the outer MW halo, where the halo in this case is assumed to exist in an inertial frame \citep{petersenReflexMotionMilky2020b}. The (near-)inertial frame referred to here is due to the very long dynamical timescales of stars in the outer halo. This assumption works well when the dynamical time of stars in the halo are much larger than the infall time of the LMC. In the inner halo of the MW however, the dynamical times are short enough that the assumption of an inertial frame from which a dipole signal is expected does not hold, as the short dynamical timescales of halo stars will also have their own response. When measuring reflex motion with respect to the outermost halo ($r>50~\rm{kpc}$), the information in the apex informs the motion of the disc due to the infall of the LMC in a psuedo-inertial frame. In the inner halo however, the response of the halo itself governs the change in apex location and magnitude with distance. In the following section, we discuss the physical response revealed by measuring the reflex motion signature in the inner halo of the MW.

\subsection{Reflex Motion of the MW disc}

\label{subsec:discussion-reflex}
The top panel of Figure~\ref{fig:cuspy-reflex} shows the magnitude of the disc velocity as a function of distance. Surprisingly, it shows little variation across the models, despite a wide coverage of MW-LMC mass ratios $ 0.09<M_{\rm LMC}/M_{\rm MW} < 0.34$. While previous works have shown that the reflex signature is sensitive to the mass ratio with fixed MW mass (see, e.g., \citealt{garavito-camargoHuntingDarkMatter2019a}, \citetalias{petersenDetectionMilkyWay2021a}, \citetalias{yaaqibRadialVariationLMCinduced2024a}), our simulations show that the MW's mass does not change the amplitude of reflex motion significantly. This suggests that the MW-LMC mass ratio may be more difficult to constrain than constraining only the LMC's mass and that the driving factor of the amplitude of the velocity dipole is the LMC's mass instead. 

The middle and lower panels of Figure~\ref{fig:cuspy-reflex} show the variation in the apex directions with the different MW haloes. Here we notice a surprisingly large variation between the models, especially in $\rm{b}_{apex}$. At large distances of $r>100~\rm{kpc}$ the dipole direction is consistent among the models, and points to a location along the past trajectory of the LMC,\footnote{Within $\sim300~\rm{kpc}$ all models have very similar trajectories (c.f. Figure~\ref{fig:cuspy-profiles}, at even smaller radii the difference in trajectories diminishes further. Since the disc's response to the LMC is strongest close to the pericentre, big differences in the apex directions at large radii are not expected.} as previously found by \citetalias{petersenDetectionMilkyWay2021a} and \citetalias{yaaqibRadialVariationLMCinduced2024a}.

\begin{figure}
    \includegraphics[width=\columnwidth]{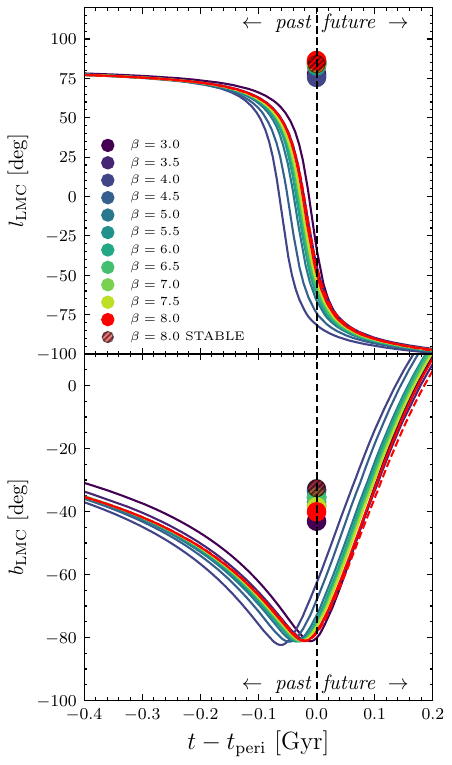}
    \caption{The past trajectory of the LMC in galactic coordinates $(l,b)$ (curves) and the apex locations $(\ell_{\rm apex}, b_{\rm apex})$ calculated at present day against all stars with $r>100~\rm{kpc}$ (points). This Figure illustrates our findings that the present-day apex points at the past location of the LMC. The present-day apex is consistent with the location of the LMC at $t-t_{\rm peri}\approx-350~\rm{Myr}$.}
    \label{fig:pasttrajLMC}
\end{figure}

Figure~\ref{fig:pasttrajLMC} shows the location of the LMC in galactic longitude and latitude with time. When compared against the apex locations at large radii (at present day), all models point at a past location of the LMC, no model points towards the present day location. From Figure~\ref{fig:pasttrajLMC}, at $t-t_{\rm peri}=-350~\rm{Myr}$ the LMC location was $\ell_{\rm LMC, past} \approx 75^\circ, b_{\rm LMC, past} \approx -40^\circ$ for all models.\footnote{A variation in $b_{\rm LMC, t=-350}$ of $\sim 5^\circ$ is present owing to the slightly different trajectories of each simulation} This is fully consistent with the convergent values of the apex locations at large radii in Figure~\ref{fig:cuspy-reflex}. The pointing of the apex locations is consistent with the interpretation of the apex location at large distances presented in \citetalias{petersenDetectionMilkyWay2021a, yaaqibRadialVariationLMCinduced2024a}.

From a dynamical standpoint, the location of the apex measured at present points at a past location of the LMC (e.g. where the LMC was $\approx300-500\rm{Myr}$ ago, of order mean dynamical timescale of the MW disc) because the displacement of the MW disc is not instantaneous, rather it happens over timescales related to the average dynamical timescales of disc stars (or the inner halo in general). Measurements of apex locations that place the direction of disc motion pointed towards the present-day location of the LMC cannot be explained by our simulations.

Furthermore, variations in the directions in the inner halo are governed by the dynamical response of the halo to the LMC's infall, which in turn affects the relative motion between the inner halo and disc (Figure~\ref{fig:cuspy-reflex}). Generally, as $\beta$ increases, the direction of $\rm{b}_{apex}$ with radius goes from being negative at $r < 40~\rm{kpc}$, then increases to a maximal value that is positive near the truncation radius where the peak value in $\rm{b}_{apex}$ increases with increasing $\beta$, then becomes negative again in the outer halo. This behaviour can be attributed to how the motion of the barycentres of the MW disc and halo move in each model.

For the most NFW-like models, the inner halo's barycentre is supported by the mass beyond the truncation radius, which impedes the barycentric motion in the inner halo induced by the LMC. In this scenario, the reflex motion signal is driven mostly by the barycentric motion of the MW disc, which leads to little change in the apex locations.

However, when truncating the halo, the barycentre of the inner halo can be moved more easily by the LMC. The motion of the inner halo in this regime is now significant, and crucially misaligned with the barycentric motion of the disc, owing to the difference in orbital timescales between halo and disc stars. This motion results in a abrupt change in measured reflex motion, this is seen in the apex directions in Figure~\ref{fig:cuspy-reflex}.


\subsection{Bulk motion of the stellar halo}
\label{subsec:discussion-bulkmotions}

\begin{figure}
    \includegraphics[width=\columnwidth]{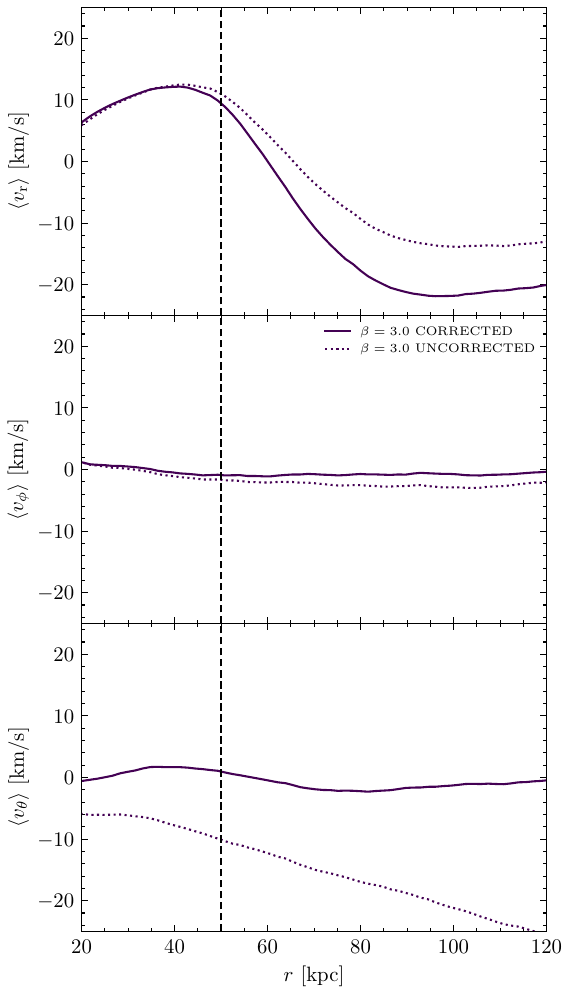}
    \caption{Similar to Figure~\ref{fig:cuspy-bulkmotions}, but highlighting the difference in the bulk motions when reflex motion is left unaccounted for in the $\beta=3.0$ model. The dotted line shows the uncorrected bulk motion signal, while the solid line shows the reflex correction signature.}
    \label{fig:reflexcorrection}
\end{figure}

As done in \citetalias{petersenDetectionMilkyWay2021a} and \citetalias{yaaqibRadialVariationLMCinduced2024a}, we measure the {\it reflex corrected} bulk motions in spherical shells about the disc barycentre at present-day in the simulations (See Figure ~\ref{fig:cuspy-bulkmotions}). We emphasise the importance of this correction, noting that if the reflex is not accounted for, the velocity dipole signal is imprinted also in the bulk motions. 

To highlight this issue, Figure~\ref{fig:reflexcorrection} shows the resulting bulk motions of the $\beta=3.0$ models in the case of accounting and not accounting for the reflex motion of the MW disc. In the case where reflex motion is unaccounted for, the velocity dipole signature imprints strongly on $\langle v_\theta\rangle$, causing the polar velocity to be an decreasing function of distance. Furthermore, it affects also the compression velocity by decreasing the magnitude of the signal at large radii. Accounting for reflex motion results in an easier interpretation of the signals - they are the kinematic response in the halo to the infall of the LMC. In \citetalias{yaaqibRadialVariationLMCinduced2024a}, the variation in the bulk motion was attributed to the presence of the bound mass of the LMC at present day, the amount of compression in the radial velocity increased with increasing LMC mass (see Figure 2 in \citetalias{yaaqibRadialVariationLMCinduced2024a}). 

In the top panel of Figure~\ref{fig:cuspy-bulkmotions}, we present the bulk motion signatures measured in each of the models shown in table \ref{tab:models}. For all models, the compression signature is clear -- beyond the break radius, the mean velocity in a given halo shell decreases with distance. However, a striking feature in our analysis shows that not only is the compression velocity present, but sinusoidal variations in $\langle v_r\rangle$ are also found. This variation occurs as truncating the MW halo induces a radial instability of breathing mode type in the halo. This is corroborated by the amplitude of $l=0$ in the basis functions for the models with modest to steep truncations in Figure~\ref{fig:coefficients}. \citet{weinbergNewDipoleInstabilities2023a} find that a truncation to the density profile between $1-2$ viral radii causes an inflection in the distribution function (a range of $E$ where $df/dE > 0$) that seeds an $l=1$ instability in the halo (See also \citealt{dattathriDipoleInstabilityGravitational2025}). Importantly, we find that when accounting for reflex motion, signatures of instability in the halo could be detected in the bulk motions. The in the outer halo, the compression velocities rebound after reaching their minima. The rebounding $\langle v_r\rangle$ shows larger peaks with increasing $\beta$ as the amplitude of the pre-existing breathing mode is larger for more truncated haloes. Furthermore, to verify that the oscillations in $\langle v_r\rangle$ are due to instability, we show in Figure \ref{fig:cuspy-bulkmotions} a stable ABG model that approximates the $\beta=8.0$ truncated model. The bulk radial motion for this model shows almost no oscillations in $\langle v_r\rangle$: $\langle v_r\rangle$ is almost constant after reaching the minimum value. The stable ABG model also shows a stable evolution in the amplitudes of the basis functions in Figure~\ref{fig:coefficients}.


In the bulk rotations ($\langle v_\phi\rangle$) of the haloes, we measure a weak signal. While this signal is small, the radius at which $\langle v_\phi\rangle$ reaches its maximum amplitude is close to the present-day radius of the LMC -- hinting that this is likely a local effect induced by the LMC.
In \citetalias{petersenDetectionMilkyWay2021a} and \citetalias{yaaqibRadialVariationLMCinduced2024a}, a prograde rotation signal of the order $40~\rm{kms^{-1}}$ was detected in the data. In those works, the rotation was attributed to primordial rotation, as the bulk motions they measure for a multitude of literature models do not have significant rotation. We arrive here at the same conclusions. Despite multiple detections of bulk rotation in the halo \citep{IorioRotation2021}, the LMC-MW interaction in simulations does not induce any significant bulk rotation in the MW halo. 

Finally, in the bottom panel of Figure~\ref{fig:cuspy-bulkmotions} we see a systematic trend in the polar motions of stars in the halo. All models have a mildly decreasing trend of $\langle v_\theta\rangle$ between $\sim20~\rm{kpc}$ and $\sim60~\rm{kpc}$ that gradually returns to zero beyond $\sim60~\rm{kpc}$. The steeper the outer halo, bigger the amplitude of $v_\theta$ found. Negative polar motion indicates that stars on average have mild motion towards the north pole of the MW. Additionally, local effects\footnote{Local effects here refers to the fact that stars within close proximity to the LMC (e.g. within the tidal radius) will be responding to the presence of the satellite at close proximity, rather than participating in the collective response of the MW halo.} could amplify the measured polar velocity signal. The breadth of the $\langle v_\theta\rangle$ signal in radius suggests that it could be a combination of local and global responses to the LMC. 


\subsection{Interpreting the reflex signals}
\label{subsec:discussion-reflexInterpretation}

We can connect the choices made for the halo profile with resulting reflex motion measured. In Figure~\ref{fig:cuspy-reflex} in the apex directions, there are two clear regimes. In the outer halo, at distances $>100~\rm{kpc}$ all the simulations point at a location along the past trajectory of the LMC. As discussed in \citetalias{petersenDetectionMilkyWay2021a} and \citetalias{yaaqibRadialVariationLMCinduced2024a}, this is due to the long dynamical timescales of stars in the outer halo, where the infall of the LMC proceeds much faster than the stars are able to respond. However, at intermediate distances, clear trends can be seen in $b_{\rm apex}$ and $\ell_{\rm apex}$. In all the simulations we run, the profiles interior to $50~\rm{kpc}$ are identical: what drives these variations at intermediate radii are the differences in the profile beyond the break radius. For the NFW halo ($\beta=3.0$) much of the mass exists in the outskirts. In this case, the inner halo can be supported gravitationally by the mass exterior to it. Therefore, during the infall of the LMC close to pericentre, the inner halo is resistant to displacement from the original barycentre. In the reflex motion signature, for the $\beta=3.0,3.5$ models, we see this effect arising as smooth and small variations (respectively) in the apex directions with distance across intermediate and large distances. 

In the case where the mass in the outskirts is removed, then the inner halo becomes more susceptible to displacement from the LMC, although the larger dynamical timescales(relative to disc stars) means that the motion of the halo will be misaligned with the motion of the disc. These two effects lead to measuring large variations at intermediate distances. To further delve deeper into the interplay between observed reflex and the halo, we notice that for the NFW halo ($\beta=3.0$), the dipole strength is the largest while the quadrupole strength is the smallest. The NFW and $\beta=3.5$ models have little variation in the apex directions. while the $\beta=8.0$ has the opposite -- small dipole and largest quadrupole at present day and a large variation in the apex direction at intermediate distances. This hints that the amount of variation in the apex locations at intermediate distances are driven by the strength of the dipole and quadrupolar deformations. 

More truncated haloes experience larger quadrupolar deformations, and are more susceptible to the influence of the LMC owing to the lack of material outwith the break radius. Thus we find that haloes that have more quadrupolar deformation will produce more variation in the apex direction.

To further investigate this, we utilise {\sc exp} to perform additional simulations of the truncated NFW $\beta=8.0$ model in Appendix~\ref{app:deformation-restriction}. Where the deformation types are restricted to having only a monopole, a monopole and dipole and a monopole, dipole and quadrupole. Figure~\ref{fig:app-restricted-deformation} shows that for a MW halo that can only deform with $l=0$, the apex locations at intermediate radii are relatively unchanged. When allowing $l=1$, $\ell_{\rm apex}$ shows a dip at $\sim50~\rm{kpc}$ and a steep rise in $b_{\rm apex}$. Allowing further the quadrupole term to contribute to the basis, $\ell_{\rm apex}$ varies more significantly at $50~\rm{kpc}$. This simple test shows that the variation in $b_{\rm apex}$ is driven by the presence of \textit{both} a monopole and a dipole, while the presence of the quadrupole enhances only the apex longitude. We discuss the implications of using simplified MW-LMC models in the Section below.

\subsection{The need for self-consistent modelling}
\label{subsec:needselfconsistent}
Rigid MW–LMC simulations have been used to study the response of the MW stellar halo to the infall of the LMC \citep[See examples of static non-deforming MW-LMC models in][]{erkalDetectionLMCinducedSloshing2021, BrooksRigidSBI2025}. In these models, the MW disc is artificially fixed at the halo barycentre, and the LMC orbits within this static configuration. A population of MW stellar halo tracer particles is then evolved under the combined potential of the MW and the infalling LMC. While such simulations have proven useful in building intuition about halo responses, they neglect several critical dynamical effects which we discuss here. 

Firstly, rigid MW models prevent reflex motion because the disc is locked at the halo centre. Recall that reflex motion is defined as the motion of the disc w.r.t to the halo \citepalias{petersenDetectionMilkyWay2021a, yaaqibRadialVariationLMCinduced2024a}. As such, these simulations cannot recover the combined observational signature of i) Disc reflex motion with respect to the halo plus ii) Internal bulk motions in the stellar halo induced by the LMC. In rigid models, bulks motion in the stellar halo resulting from the LMC interaction are interpreted as reflex motion, conflating with our definition and leading to confusion in the interpretation of the measured kinematics.

Rigid models do not allow LMC-induced deformation of the MW and LMC DM haloes. For example, these models do not account for the formation of a DM wake trailing the LMC, which in turn affects the dynamics of halo stars.
Using self-consistent $n$-body simulation we demonstrate in Appendix~\ref{app:deformation-restriction} that suppressing higher order ($l>0$) halo deformation erases the dipole and quadrupole contributions to the radial variation of the apex locations and disc velocity.  Moreover, while rigid setups can prescribe dynamical friction in the trajectory calculation, they cannot model LMC mass loss or the subsequent mass deposition in the MW halo. They also fail to capture instabilities if any are present (See Section~\ref{subsec:discussion-bulkmotions}).

Additionally rigid models have been used to model the effect of the LMC on stellar streams \citep{erkalTotalMassLarge2019a,koposovS5ProbingMilky2023}, however \citet{lilleengenEffectDeformingDark2022} has shown that higher order deformations matter for precise matching of the simulations with streams.

We therefore caution against interpreting observational data using rigid MW–LMC models and stress the importance of using self-consistent $n$-body simulations that simultaneously model the coupled dynamics of the disc, halo and LMC.\footnote{We note that while we use a rigid (non-deforming) LMC model in this work, the MW itself is fully self-consistent. This simplification was adopted as our work concerns mainly the response of the halo to the infall of the LMC.}

\section{Conclusions}
\label{sec:Conclusion}
In this work, we presented a suite of MW-LMC simulations varying the outer density slope of the MW halo between $3.0 < \beta < 8.0$. We self-consistently simulated the infall of the LMC and extracted for each simulation the reflex motion parameters $v_{\rm travel}, \ell_{\rm apex}, b_{\rm apex}, \langle v_r\rangle, \langle v_\phi\rangle, \langle v_\theta\rangle$ as a function of Galactocentric distance at present day. We also show the deformations experienced in each simulation through the amplitude of basis function coefficients as a function of time for the monopole, dipole and quadrupolar deformations. We summarise our findings as follows:

\begin{enumerate}
    \item At fixed LMC mass, the amplitude of reflex motion $v_{\rm travel}$ does not vary significantly for a wide range of MW-LMC mass ratios of $0.15 < M_{\rm LMC}/M_{\rm MW} < 0.35$. Our results imply that $v_{\rm travel}$ depends on the assumed mass of the LMC, rather than the mass ratio of the MW-LMC system. This result cautions against our ability to constrain the MW-LMC mass ratio through the detection of the velocity dipole. 
    \item The direction of the velocity dipole at intermediate radii between $30-60~{\rm kpc}$ can vary based on the {\it outer} MW halo profile. Our results show that (a) steeper truncations to the outer MW halo produce apex directions that vary more strongly with radius (b) the apex locations when measured against the outer halo is not sensitive to the outer MW halo profile, but to the trajectory of LMC. This may be due to the fact that the inner haloes of more steeply truncated models are less gravitationally supported by DM mass in the outskirts. 
As a result, the inner halo responds more strongly to the LMC's infall, although not with the same timescales as the disc. This leads to a stronger misalignment between the motion of the disc and inner halo, and therefore a larger variation in the apex locations.
    \item We show also explicitly that large deviations in the apex locations at intermediate radii are caused by the dipolar and quadrupolar deformations to the halo, through simulations that restrict the type of deformations allowed in the halo (See Appendix~\ref{app:deformation-restriction}).
    \item In the compression velocity of the halo, we show that all MW models show a clear compression signal, but the maximal compression experienced by the models is strongest for the shallow truncations and weakest for more truncated halos. 
    \item We discover a radial (breathing mode) instability in truncated MW halos that forms as a result of the model truncation. We find that the amplitude of the oscillations increases with increasing outer slope. We find that this instability presents in the compression velocity $\langle v_r\rangle$ as oscillations that vary with distance.
\end{enumerate}

We conclude with the following remarks. The reflex motion signature holds information on the outer profile of the MW DM halo.
While we find no empirical relation between the outer halo slope and reflex motion parameters, our experiment shed light on the dynamical relations between the inner MW halo, outer MW halo and MW disc to the infall of the LMC. 

Furthermore, through analysis of the basis function expansion coefficients in time, we show that in order to measure a $\vec{v}_{\rm travel}$ that varies in direction at intermediate radii, the MW halo must deform with at least $l=1$, and in order to measure compression in $\langle v_r \rangle$ an $l=0$ deformation is needed. Both the apex directions and compression have been measured to be radially varying in the literature. 

We stress the need to account for reflex motion when computing the bulk velocities in the MW halo at all radii, and show that when accounting for reflex instabilities in the halo could be measured for large-amplitude breathing modes. 
We further stress the need for self-consistent models of the MW-LMC interaction in Section~\ref{subsec:needselfconsistent}. We emphasise that rigid MW-LMC do not account for deformations of the halo, which we show have a significant impact of the resulting reflex motion signature (See Appendix~\ref{app:deformation-restriction}).

Future work could see a more complex reflex motion model that accounts for cyclic variations in the compression velocity, to test whether a radial instability exists in the MW halo. Additionally, further testing with a different family of models (e.g. using stable ABG models varying only the transition $\alpha$) could provide important insights about the pre-LMC infall shape of the MW halo.
\section*{Acknowledgements}

MSP acknowledges funding from a UKRI Stephen Hawking Fellowship. {\sc exp} is maintained by the {\sc exp} collaboration (\hyperlink{https://github.com/exp-code}{https://github.com/EXP-code}).  We acknowledge and thank the developers of the following software that was used in this work: NumPy~\citep{harrisArrayProgrammingNumPy2020}, SciPy~\citep{virtanenSciPy10Fundamental2020}, IPython~\citep{perezIPythonSystemInteractive2007}, Matplotlib~\citep{hunterMatplotlib2DGraphics2007}, Jupyter~\citep{communityJupyterBook2021a}.
    
\section*{Data Availability}
All the simulations snapshots used in this work, and analysis files are available from the corresponding author upon reasonable request.


\bibliographystyle{mnras}
\bibliography{references}


\appendix
\section{Deformation restricted simulations}
\label{app:deformation-restriction}
\begin{figure}
    \centering
    \includegraphics[width=\columnwidth]{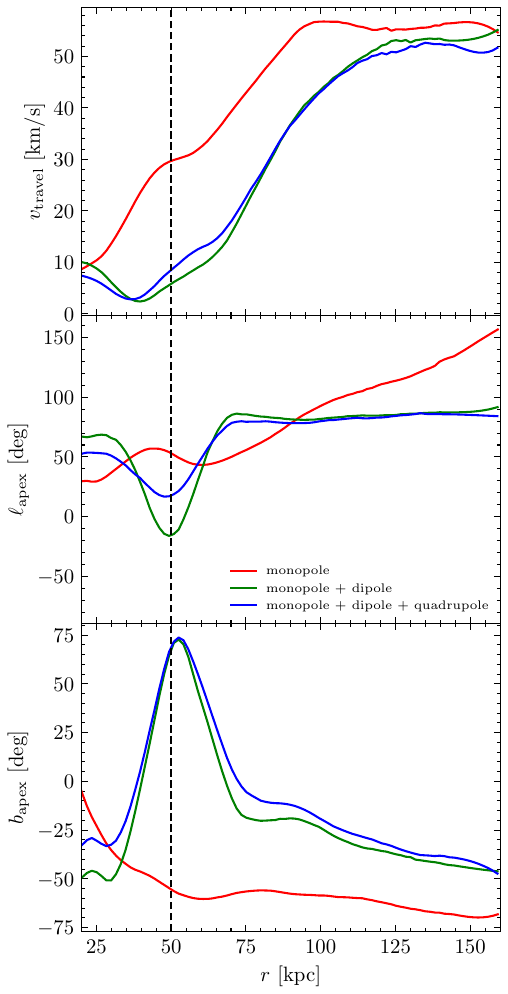}
    \caption{This figure shows the present day disc travel velocity(top panel), apex longitude(middle panel) and apex latitude (bottom panel) in three simulations of the $\beta=8.0$ model. Where the type of deformations was restricted to monopole only (red line), monopole and dipole (green line) and monopole, dipole and quadrupole (blue line). The dip in $\ell_{\rm apex}$ and steep rise in $b_{\rm apex}$ are only present when higher order deformations are allowed. The monopole-only simulations shows little variations in the apex directions at intermediate radii.}
    \label{fig:app-restricted-deformation}
\end{figure}


In this Appendix we simulate the $\beta=8.0$ truncated NFW in {\sc exp} but restrict the type of deformations that the MW to three scenarios - monopole only, monopole and dipole, monopole, dipole and quadrupole. In the software this is done by running the simulations setting $l_{\rm max}$ to either 0, 1 and 2 respectively (See Table~\ref{tab:models} for the values of $l_{\rm max},n_{\rm max}$ used in the main text).
These simulations were run to investigate which deformation type cause changes in the apex locations found for most models in Section~\ref{sec:results} at $\sim50~\rm{kpc}$, connecting density deformations to MW reflex. The results of these simulations are shown in Figure \ref{fig:app-restricted-deformation}. We find that for the monopole only simulation (red curve in figure) there is no significant change to the apex locations at intermediate radii. However, when a dipole deformation is allowed (green curve), $b_{\rm apex}$ shows a steep rise around $50~\rm{kpc}$ followed by a falling after reaching the maximal value and flattening at large radii. No significant changes in the profile of $b_{\rm apex}$ are found between $l_{\rm max}=1$ and $l_{\rm max}=2$ simulations. In $\ell_{\rm apex}$, MW halo deformations with $l_{\rm max}>0$ both lead to a dip in the profile of the apex longitude at intermediate distances. 

Our simple test here shows two main insights: i) the dipole (and monopole) drives changes in the apex locations in our simulations about $50~\rm{kpc}$ ii) allowing a quadrupolar deformation (along with monopole and dipole) results in changes to only the apex longitude.


\bsp	
\label{lastpage}
\end{document}